\documentclass[12pt,a4paper]{article}

\usepackage{amsthm,amsmath,natbib,amssymb,amsfonts,bm}
\usepackage{graphicx}
\usepackage{placeins}

\def\hang{\hangindent\parindent}
\def\rf{\par\noindent\hang}

\newtheorem*{theorem*}{Theorem}

\theoremstyle{definition}

\theoremstyle{remark}

\newcommand{\bx}{\boldsymbol{x}}
\newcommand{\bz}{\boldsymbol{z}}
\newcommand{\bh}{\boldsymbol{h}}
\newcommand{\bc}{\boldsymbol{c}}
\newcommand{\bu}{\boldsymbol{u}}
\newcommand{\bk}{\boldsymbol{k}}
\newcommand{\bl}{\boldsymbol{l}}
\newcommand{\bv}{\boldsymbol{v}}
\newcommand{\bzero}{\boldsymbol{0}}

\DeclareMathAlphabet{\mathpzc}{OT1}{pzc}{m}{it}

\begin{document}

\baselineskip=21pt

\phantom{1}
\vspace{-2cm}
\begin{center}
 {\bf \Large A new method of randomization of lattice rules for multiple integration}
\end{center}

\bigskip

\begin{center}
\large{{\bf Paul Kabaila$^*$}}
\end{center}

\begin{center}
{\sl Department of Mathematics and Statistics,
La Trobe University, Victoria 3086, Australia}
\end{center}


\noindent {\bf Abstract}

\medskip

Cranley and Patterson put forward the following randomization as the basis for the estimation of the error
of a lattice rule for an integral of a one-periodic function over the unit cube in $s$ dimensions.
The lattice rule is randomized
using independent random shifts in each coordinate direction that are
uniformly distributed in the interval $[0,1]$.
This randomized lattice rule results in an unbiased estimator of the multiple integral.
However, in practice,
random variables that are independent and uniformly distributed on $[0,1]$
are not available, since this would require an infinite number of random
independent bits.
A more realistic practical implementation of the Cranley and Patterson randomization uses
$r s$ independent random bits, in the following way.
The lattice rule is randomized using independent random shifts in
each coordinate direction that are uniformly distributed
on $\big \{0, 1/2^r, \dots, (2^r-1)/2^r \big \}$, where $r$ may be large.
For a rank-1 lattice
rule with $2^m$ quadrature points and $r \ge m$, we show that this randomized lattice
rule leads to an estimator of the multiple integral that typically has a large bias.
We therefore propose that these $r s$ independent random bits be used to perform a new randomization
that employs an extension, in the number of quadrature points,
to a lattice rule with $2^{m+sr}$ quadrature points (leading to embedded lattice rules).
This new randomization is shown to lead to an estimator
of the multiple integral that has much smaller bias.

\bigskip


\vspace{0.7cm}

\noindent $^*$Corresponding author. Tel.: +61 3 9479 2594, fax: +61 3 9479 2466.

\noindent {\sl E-mail address:} P.Kabaila@latrobe.edu.au (Paul Kabaila).

\newpage

\baselineskip=20pt

\noindent {\large\textbf{1. Introduction}}

\medskip

\noindent In this paper we consider the problem of computing
\begin{equation*}
If = \int_{C^s} f(\bx) \, d \bx,
\end{equation*}
where $f: \mathbb{R}^s \rightarrow \mathbb{R}$, $C^s = [0,1]^s$ and the function
$f$ is one-periodic with respect to each component of $\bx$, i.e. $f(\bx) = f(\bx+\bz)$
for all $\bz \in \mathbb{Z}^s$ and $\bx \in \mathbb{R}^s$.
We suppose that $f$ has an absolutely convergent Fourier series representation
\begin{equation*}
f(\bx) = \sum_{\bh \in \mathbb{Z}^s} \hat{f}(\bh) \, e^{2 \pi i \bh \cdot \bx},
\end{equation*}
where $\bh \cdot \bx = h_1 x_1 + \dots + h_s x_s$ is the inner product in $\mathbb{R}^s$.
We suppose further that $f$ is known to belong to some class of functions $F$ of
smooth functions, with smoothness measured by the rate of decay of the Fourier
coefficients. For classes of sufficiently smooth functions, a remarkably accurate
approximation to $If$ is provided by a lattice rule
\begin{equation*}
Qf = \frac{1}{N} \sum_{j=0}^{N-1} f(\bx_j),
\end{equation*}
where $\big \{ \bx_0, \dots, \bx_{N-1} \big \}$ are the points of a carefully chosen
integration lattice $L \subset \mathbb{R}^s$ that lie in the half-open cube $[0,1)^s$.
An integration lattice in $\mathbb{R}^s$ is defined as a discrete subset of $\mathbb{R}^s$
which is closed under addition and subtraction and which contains $\mathbb{Z}^s$ as a
subset. A very readable introduction to lattice rules is provided by Sloan and Joe (1994).

The standard method for estimating the lattice rule error $Qf - If$ is the randomization
method due to Cranley and Patterson (1976). Define the shifted lattice rule
\begin{equation*}
Q_c f = \frac{1}{N} \sum_{j=0}^{N-1} f(\{\bx_j + \bc\})
\end{equation*}
where $\bc \in \mathbb{R}^s$ and $\{ \bx \}$ denotes the fractional part of the vector $\bx$,
obtained by taking the fractional part of each component of $\bx$. Theorem 2.10 of
Sloan and Joe (1994) states that
\begin{equation}
\label{error_shifted_lattice}
Q_{\bc} f - I f = \mathop{{\sum}'}_{\bh \in L^{\bot}} \, e^{2 \pi i \bh \cdot \bc} \, \hat{f}(\bh),
\end{equation}
where the prime on the sum indicated that the zero term is omitted from the sum and
$L^{\bot}$ denotes the dual lattice, defined e.g. on p.26 of Sloan and Joe (1994).

Let $y_1 = Q_{\bu_1} f, \dots, y_q = Q_{\bu_q} f$, where $\bu_1, \dots, \bu_q$ are independent and
identically distributed (iid) uniformly in the unit cube $[0,1]^s$.
Note that $\bu$ being uniformly distributed in $[0,1]^s$ is equivalent to the components of
$\bu$ being iid uniformly distributed on $[0,1]$. Cranley and Patterson (1976) propose
that $I f$ be estimated by
\begin{equation*}
\bar{y} = \frac{1}{q} \sum_{k=1}^q y_k.
\end{equation*}
As is well-known, $E(y_k) = I f$ for $k=1, \dots, q$, so that $E(\bar{y}) = I f$. In other words,
$y_k$ is an unbiased estimator of $I f$ for $k=1, \dots, q$, so that $\bar{y}$ is an unbiased estimator
of $I f$.
An expression for Var($\bar{y}$),
in terms of the Fourier coefficients of $f$, is provided by Proposition 4 of L'Ecuyer and Lumieux (2000).
We review these known results in the Appendix and show that expressions for
higher-order moments can also be found in terms of the Fourier coefficients of $f$.

The theory for the randomization proposed by Cranley and Patterson (1976) is elegant and relatively
simple. However, this form of randomization is an idealisation.
In practice, observations of
random variables that are independent and uniformly distributed on $[0,1]$
are not available. Henceforth, when we refer to iid random bits $b_1, b_2, \dots$,
we assume that $P(b_1 = 0) = P(b_1 = 1) = 1/2$.
Even a single random variable uniformly distributed on $[0,1]$,
is equivalent to an infinite number of iid
random bits $b_1, b_2, \ldots$.
Specifically, if $u$ is uniformly distributed on $[0,1]$ then $u = .b_1 b_2 \cdots$.

If one replaces these independent and uniformly distributed (on $[0,1]$)
random variables
by pseudorandom numbers then the resulting estimator of the multiple integral will have largely
unpredictable properties. What properties might be rigorously established would be hard won and of
limited scope and would come nowhere near the elegance and simplicity of the theory for the randomization
put forward by Cranley and Patterson (1976).

Observations of truly random bits may be taken from physical sources such as electronic
thermal noise and various sources that rely on quantum-mechanical effects.
Such observations may be obtained through the internet from reputable sources such as
the {\sl ANU Quantum Random Numbers Server} (http://qrng.anu.edu.au). Of course, in practice, one can obtain
observations of only a finite number of random bits.

A realistic practical implementation of the Cranley and Patterson randomization uses
$r s$ independent random bits, in the following way.
The lattice rule is randomized using independent random shifts in
each coordinate direction that are uniformly distributed
on $\big \{0, 1/2^r, \dots, (2^r-1)/2^r \big \}$, where $r$ may be large.

For simplicity of exposition, we consider the particular case that the integration lattice
is generated by a rank-1 lattice rule
\begin{equation}
\label{rank_1_rule}
Qf = \frac{1}{N} \sum_{j=0}^{N-1} f \left( \left \{ \frac{j}{N} \bz \right \} \right ),
\end{equation}
where the number of quadrature points $N=2^m$, $\bz \in \mathbb{Z}^s$ and has no common factors with $N$.
We call $\bz$ the {\sl generating vector}.
For the definition of the rank of a lattice rule see e.g. Sloan and Joe (1994).
Rank-1 lattice rules
were introduced and extensively analysed by Korobov (1959). These rules may be found
using the component-by-component (CBC) construction (see e.g. Dick, Kuo and Sloan, 2014).
Also, for simplicity, we suppose that $r \ge m$.
In Section 2, we show that this
randomization leads to an estimator of the multiple integral that typically has a large bias.
Note that $s$ random variables that are iid and uniformly distributed
on $\big \{0, 1/2^r, \dots, (2^r-1)/2^r \big \}$ can be transformed to $rs$ iid random bits
$b_1, \dots, b_{rs}$ and vice versa.

We therefore propose, in Section 3, that these $r s$ iid random bits be used to perform a new randomization
that employs an extension, in the number of quadrature points,
from a rank-1 lattice rule with $2^m$ quadrature points
to a rank-1 lattice rule with $2^{m+sr}$ quadrature points.
This new randomization is shown to lead to an estimator
of the multiple integral $I f$ that has much smaller bias.
Some numerical illustrations of this property
are provided in Section 4.

\bigskip

\noindent {\large\textbf{2. Results for randomization (of the type proposed by Cranley and Patterson, 1976)
using iid random variables uniformly distributed on
$\boldsymbol{ \big \{0, 1/2^r, \ldots, (2^r-1)/2^r \big \} }$ }}

\medskip

Consider the rank-1 lattice rule \eqref{rank_1_rule}, where $N=2^m$. The
randomization of this lattice rule proposed by Cranley and Patterson (1976) is
\begin{equation}
\label{rank_1_lattice_randomized_U[0,1]}
Q_{\bu} f = \frac{1}{N} \sum_{j=0}^{N-1} f \left( \left \{ \frac{j}{N} \bz + \bu \right \} \right ),
\end{equation}
where the random vector $\bu$ is uniformly distributed on $[0,1]^s$. As already noted in the introduction,
it is unrealistic to assume that we have observations of the infinite number of truly random bits
that are required to produce an observation of $\bu$. The randomized lattice rule that corresponds
to \eqref{rank_1_lattice_randomized_U[0,1]} and is based on only a finite number $r s$ of random bits
is
\begin{equation*}
Q_{\bv} f = \frac{1}{N} \sum_{j=0}^{N-1} f \left( \left \{ \frac{j}{N} \bz + \bv \right \} \right ),
\end{equation*}
where the random vector $\bv$ is uniformly distributed on $\{0, 1/2^r, \dots, (2^r-1)/2^r\}^s$,
where $r$ may be large.
For simplicity, we assume that $r \ge m$. In this section, we show that the bias
$E(Q_{\bv} f) - I f$ can be unacceptably large for $f$ in the class $F$.

To find $E(Q_{\bv} f)$, we proceed as follows. The $k$'th component of the random $s$-vector
\begin{equation}
\label{random_s_vector}
\left \{ \frac{j}{N} \bz + \bv \right \} = \left \{ \frac{j}{2^m} \bz + \bv \right \}
\end{equation}
is
\begin{equation}
\label{kth_component_of_random_s_vector}
\left \{ \frac{j}{2^m} z_k + v_k \right \},
\end{equation}
where $z_k$ and $v_k$ denote the $k$'th components of $\bz$ and $\bv$, respectively.
Since we have assumed that $r \ge m$, \eqref{kth_component_of_random_s_vector}
is uniformly distributed on $\{0, 1/2^r, \dots, (2^r-1)/2^r\}$.
Also, the components of the random $s$-vector \eqref{random_s_vector}
are independent random variables since the components of $\bv = (v_1, \dots, v_s)$
are independent. Thus \eqref{random_s_vector} has the same probability distribution as $\bv$.

Hence
\begin{align*}
E(Q_{\bv} f)
&= E \left [ \frac{1}{N} \sum_{j=0}^{N-1} f \left( \left \{ \frac{j}{N} \bz + \bv \right \} \right ) \right ] \\
&=  \frac{1}{N} \sum_{j=0}^{N-1} E \left [ f \left( \left \{ \frac{j}{N} \bz + \bv \right \} \right ) \right ] \\
&=  \frac{1}{N} \sum_{j=0}^{N-1} E \left [ f (\bv) \right ] \\
&=  E \left [ f (\bv) \right ] \\
&= \frac{1}{2^{sr}}
\sum_{j_1=0}^{2^r - 1} \dots \sum_{j_s=0}^{2^r - 1}
f \left ( \left ( \frac{j_1}{2^r}, \dots, \frac{j_s}{2^r} \right )  \right ).
\end{align*}
This is just the product-rectangle rule with $2^{sr}$ quadrature points.
This rule is known to be a particularly bad lattice rule for the types of function classes $F$
under consideration. Furthermore, the number of quadrature points in only $2^{sr}$, functionally
independent of $m$. Therefore, the magnitude of the bias $|E(Q_v f) - I f|$ will take unacceptably large values
for some $f$'s in the
class of functions $F$. This is a very serious disadvantage of this form of randomization.

\newpage


\noindent {\large\textbf{3. New randomization using iid random variables uniformly distributed on
$\boldsymbol{ \big \{0, 1/2^r, \ldots, (2^r-1)/2^r \big \} }$ and two embedded lattice rules}}

\medskip

Note that $s$ random variables that are iid and uniformly distributed
on \newline $\big \{0, 1/2^r, \dots, (2^r-1)/2^r \big \}$ can be transformed to $rs$ iid random bits
$b_1, \dots, b_{rs}$ and vice versa. In this section, we describe a new randomization that uses these
random bits to randomize the rank-1 lattice rule \eqref{rank_1_rule}, which has $N=2^m$
quadrature points. This randomization has far better properties than the randomization
described in the previous section that uses the same number of iid random bits.

We choose the generating vector $\bz$ such that both the rank-1 lattice rule
\begin{equation}
\label{rank_1_lattice_rule}
Qf = \frac{1}{2^m} \sum_{j=0}^{2^m-1} f \left( \left \{ \frac{j}{2^m} \bz \right \} \right )
\end{equation}
and the rank-1 lattice rule
\begin{equation}
\label{extended_rank_1_lattice_rule}
\tilde{Q}f = \frac{1}{2^{m+sr}} \sum_{k=0}^{2^{m+sr}-1} f \left( \left \{ \frac{k}{2^{m+sr}} \bz \right \} \right )
\end{equation}
perform well in the class $F$ of functions $f$. This generating vector can be found using the method
of Cools, Kuo and Nuyens (2006). The quadrature points of $Qf$ are embedded in the set of quadrature
points of $\tilde{Q}f$. Extensions of lattice rules in both dimension $s$ and number of quadrature
points were introduced by Hickernell, Hong, L'Ecuyer and Lemieux (2000). An existence proof for
good extensible rank-1 lattice rules is provided by Hickernell and Niederreiter (2003).

Let $w$ be a random variable that is uniformly distributed on $\{0, 1/2^{sr}, \dots, (2^{sr}-1)/2^{sr} \}$.
We can express $w$ in binary form as $w = .b_1 \dots b_{sr}$, where $b_1, \dots, b_{sr}$ are iid random
bits. The new randomized rank-1 lattice rule is
\begin{equation*}
Q^{\prime}_w f = \frac{1}{2^m} \sum_{j=0}^{2^m-1} f \left( \left \{ \frac{j+w}{2^m} \, \bz \right \} \right )
\end{equation*}
It may be shown that $E[Q^{\prime}_w f] = \tilde{Q} f$ as follows.
\begin{align*}
E[Q^{\prime}_w f] &= \frac{1}{2^m} \sum_{j=0}^{2^m-1} E \left[f \left( \left \{ \frac{j+w}{2^m} \, \bz \right \} \right ) \right] \\
&= \frac{1}{2^m} \sum_{j=0}^{2^m-1} \frac{1}{2^{sr}}
\sum_{\ell=0}^{2^{sr}-1} f \left( \left \{ \frac{j+(\ell/2^{sr})}{2^m} \, \bz \right \} \right ) \\
&= \frac{1}{2^{m+sr}} \sum_{j=0}^{2^m-1}
\sum_{\ell=0}^{2^{sr}-1} f \left( \left \{ \frac{2^{sr} j+\ell}{2^{m+sr}} \, \bz \right \} \right ) \\
&= \frac{1}{2^{m+sr}} \sum_{k=0}^{2^{m+sr}-1}
f \left( \left \{ \frac{k}{2^{m+sr}} \, \bz \right \} \right ),
\end{align*}
since $2^{sr} j+\ell$ can be expressed as a binary number using $m+sr$ bits.

Let $y_1 = Q^{\prime}_{w_1} f, \dots, y_q = Q^{\prime}_{w_q} f$, where $w_1,\dots, w_q$ are iid uniformly
distributed on $\{0, 1/2^{sr}, \dots, (2^{sr}-1)/2^{sr} \}$. We estimate $I f$ by
\begin{equation*}
\bar{y} = \frac{1}{q} \sum_{k=1}^q y_k.
\end{equation*}
Obviously, $E(\bar{y}) = \tilde{Q} f$. Therefore, $\bar{y}$ is a biased estimator of $I f$. However,
the way in which the generating vector $\bz$ has been chosen implies that the bias
$E(\bar{y}) - I f = \tilde{Q} f - I f$ will be small.

So far, we have randomized using iid random bits because this is the way that observations
of truly random variables usually present themselves in practice. We remark that it is straightforward
to develop corresponding results for randomizations using
iid random variables that are uniformly distributed
on $\{0, \dots, b-1\}$ for some base $b$ other than $b=2$.

\bigskip

\noindent {\large\textbf{4. Numerical illustrations}}

\medskip

To numerically illustrate the much smaller bias of the estimator that results from the new randomization described
in the previous section, we consider the numerical integration of a function $f$ whose integral $I f$ takes a known
value. A convenient class of such functions, which has been used extensively in the construction of lattice rules
with good properties, is described on p. 72--73 of Sloan and Joe (1994). The particular member of this of this class
that we consider is the following. For
$\bx = (x_1, \dots, x_s)$, we suppose that
\begin{equation*}
f(\bx) = \prod_{i=1}^s \big( 1 + B_2(x_i)  \big),
\end{equation*}
where $B_2(x) = x^2 - x + (1/6)$. Obviously, $I f = 1$.

We consider rank-lattice rules, of the form first proposed by Korobov, for which the
generating vector
\begin{equation*}
\bz = \left( 1 , \ell, \ell^2, \dots, \ell^{s-1} \right),
\end{equation*}
where $\ell$ is a carefully chosen positive integer. In particular, we consider the following three values of $\ell$:
17797, 1267 and 12915. These values are taken from Table 4.1 of
Hickernell, Hong, L'Ecuyer and Lemieux (2000).
We also consider the following two values of $(s,m,r)$: $(s,m,r) = (3,4,4)$
and $(s,m,r) = (2,5,5)$. It is expected that both rank-1 lattice rules
\eqref{rank_1_lattice_rule}
and
\eqref{extended_rank_1_lattice_rule}
will perform well for both of these values of $(s,m,r)$.
It is therefore expected that the new randomization method, described in Section 3, will lead to the estimator
$Q^{\prime}_w f$ of the multiple integral $I f$ that has much smaller bias that the estimator $Q_{\bv} f$ of this multiple integral that results from the
randomization (of the type proposed by Cranley and Patterson, 1976)
using iid random variables uniformly distributed on
$\big \{0, 1/2^r, \ldots, (2^r-1)/2^r \big \}$, described in Section 2.
This expectation is borne out by the numerical results presented in Tables 1 and 2.

\bigskip

\begin{table}[h]
\begin{center}
\begin{tabular}{|c|c|c|c|}
\hline
  Bias &  $\ell = 17797$ & $\ell = 1267$ \phantom{1} & $\ell = 12915$ \\
  \hline
  $E(Q_{\bv} f) - 1$ &  $1.9544 \times 10^{-3}$ & $1.9544 \times 10^{-3}$ &  $1.9544 \times 10^{-3}$ \\
  \hline
  $E(Q^{\prime}_w f) - 1$ &  $5.1619 \times 10^{-9}$ & $1.5158 \times 10^{-8}$ &  $1.9155 \times 10^{-8}$ \\
  \hline
\end{tabular}
\end{center}
\caption{Results for $(s,m,r) = (3,4,4)$. Comparison of the bias $E(Q_{\bv} f) - 1$ of the estimator that results from the
randomization (of the type proposed by Cranley and Patterson, 1976)
using $s r$ iid random bits with the bias $E(Q^{\prime}_w f) - 1$ of the estimator that results
from the new randomization, which also uses $s r$ iid random bits.}
\end{table}

\bigskip

\begin{table}[h]
\begin{center}
\begin{tabular}{|c|c|c|c|}
\hline
  Bias &  $\ell = 17797$ & $\ell = 1267$ \phantom{1} & $\ell = 12915$ \\
  \hline
  $E(Q_{\bv} f) - 1$ &  $3.2555 \times 10^{-4}$ & $3.2555 \times 10^{-4}$ & $3.2555 \times 10^{-4}$ \\
  \hline
  $E(Q^{\prime}_w f) - 1$ &  $1.2940 \times 10^{-9}$ & $4.4993 \times 10^{-9}$ & $1.7820 \times 10^{-9}$ \\
  \hline
\end{tabular}
\end{center}
\caption{Results for $(s,m,r) = (2,5,5)$. Comparison of the bias $E(Q_{\bv} f) - 1$ of the estimator that results from the
randomization (of the type proposed by Cranley and Patterson, 1976)
using $s r$ iid random bits with the bias $E(Q^{\prime}_w f) - 1$ of the estimator that results
from the new randomization, which also uses $s r$ iid random bits.}
\end{table}

It is also of interest to compare the standard deviations of the estimators $Q_{\bv} f$
and $Q^{\prime}_w f$. These standard deviations are the square roots of
\begin{equation*}
\text{Var}(Q_{\bv} f) = \frac{1}{2^{r s}} \sum_{\bv \in \{0, 1/2^r, \dots, (2^r - 1)/2^r \}^s}
\big( Q_{\bv} f - E(Q_{\bv} f) \big)^2
\end{equation*}
and
\begin{equation*}
\text{Var}(Q^{\prime}_w f) = \frac{1}{2^{r s}} \sum_{w \in \{0, 1/2^{sr}, \dots, (2^{sr} - 1)/2^{sr} \}}
\big( Q^{\prime}_w f - E(Q^{\prime}_w f) \big)^2,
\end{equation*}
respectively. Some numerical values for these standard deviations are presented in Tables 3 and 4.
These tables show that, for each of the cases considered, (a) these standard deviations are close and (b) the standard
deviation of the estimator $Q_{\bv} f$ is comparable to the magnitude of its bias.

\bigskip

\begin{table}[h]
\begin{center}
\begin{tabular}{|c|c|c|c|}
\hline
  Standard deviation &  $\ell = 17797$ & $\ell = 1267$ \phantom{1} & $\ell = 12915$ \\
  \hline
  $\sqrt{\text{Var}(Q_{\bv} f)}$ &  $7.938 \times 10^{-4}$ & $7.938 \times 10^{-4}$ &  $7.938 \times 10^{-4}$ \\
  \hline
  $\sqrt{\text{Var}Q^{\prime}_w f)}$ &  $8.389 \times 10^{-4}$ & $8.374 \times 10^{-4}$ &  $8.378 \times 10^{-4}$ \\
  \hline
\end{tabular}
\end{center}
\caption{Results for $(s,m,r) = (3,4,4)$. Comparison of the standard deviation of the estimator $Q_{\bv} f$ that results from the
randomization (of the type proposed by Cranley and Patterson, 1976)
using $s r$ iid random bits with the standard deviation of the estimator $Q^{\prime}_w f$ that results
from the new randomization, which also uses $s r$ iid random bits.}
\end{table}

\bigskip

\begin{table}[h]
\begin{center}
\begin{tabular}{|c|c|c|c|}
\hline
  Standard deviation &  $\ell = 17797$ & $\ell = 1267$ \phantom{1} & $\ell = 12915$ \\
  \hline
  $\sqrt{\text{Var}(Q_{\bv} f)}$ &  $1.6598 \times 10^{-4}$ & $1.6598 \times 10^{-4}$ & $1.6598 \times 10^{-4}$ \\
  \hline
  $\sqrt{\text{Var}Q^{\prime}_w f)}$ &  $1.8194 \times 10^{-4}$ & $1.820\times 10^{-4}$ & $1.782 \times 10^{-4}$ \\
  \hline
\end{tabular}
\end{center}
\caption{Results for $(s,m,r) = (2,5,5)$. Comparison of the standard deviation of the estimator $Q_{\bv} f$ that results from the
randomization (of the type proposed by Cranley and Patterson, 1976)
using $s r$ iid random bits with the standard deviation of the estimator $Q^{\prime}_w f$ that results
from the new randomization, which also uses $s r$ iid random bits.}
\end{table}

\bigskip

\noindent {\large\textbf{5. Conclusion}}

\medskip

The new randomization method described in Section 3 requires the extension of a rank-1 lattice
rule, for given $s$, in the number of quadrature points
from $2^m$ to $2^{m + sr}$, where $s r$ may be relatively large.
Finding such an extension that leads to embedded lattice rules with very good properties for both
$2^m$ quadrature points and $2^{m + sr}$ quadrature points is clearly an important task.

\bigskip

\noindent {\bf Acknowledgment}

\medskip

The author is grateful to Josef Dick, Frances Kuo and Pierre L'Ecuyer for helpful discussions.

\bigskip

\noindent {\bf References}

\smallskip

\rf Barndorff-Nielsen, O.E. and Cox, D.R. (1989). {\sl Asymptotic Techniques for Use in Statistics}.
Chapman and Hall, London.

\smallskip



\rf Cools, R., Kuo, F. and Nuyens, D. (2006). Constructing embedded lattice rules for
multivariate integration. {\sl SIAM Journal of Scientific Computing}, 28, 2162--2188.

\smallskip

\rf Cranley, R. and Patterson, T.N.L. (1976). Randomization of number theoretic methods
for multiple integration. {\sl SIAM Journal of Numerical Analysis}, 13, 904--914.



\smallskip



\rf Hickernell, F.J., Hong, H.S., L'Ecuyer, P. and Lemieux, C. (2000). Extensible lattice
sequences for quasi-Monte Carlo quadrature. {\sl SIAM Journal of Scientific Computing}, 22, 1117--1138.

\smallskip

\rf Hickernell, F.J. and Niederreiter, H. (2003). The existence of good extensible rank-1 lattice
rules. {\sl Journal of Complexity}, 19, 286--300.

\smallskip

\rf Korobov, N.M. (1959). The approximate computation of multiple integrals (in
Russian). {\sl Dokl. Akad. Nauk SSSR}, 124, 1207–-1210.

\smallskip

\rf L'Ecuyer, P., Munger, D. and Tuffin, B. (2010). On the distribution of integration error
for randomly-shifted lattice rules. {\sl Electronic Journal of Statistics}, 4, 950--993.

\smallskip

\rf L'Ecuyer, P. and Lemieux, C. (2000). Variance reduction by lattice rules.
{\sl Management Science}, 46, 1214--1235.

\smallskip

\rf Sloan, I.H. and Joe, S. (1994). {\sl Lattice Methods for Multiple Integration}. Clarendon Press, Oxford.

\newpage


\noindent {\large\textbf{Appendix: Review and extension of results for the randomization
of Cranley and Patterson (1976) using iid
random variables uniformly distributed on $\boldsymbol{[0,1]}$}}

\medskip

Let $y_1 = Q_{\bu_1} f, \dots, y_q = Q_{\bu_q} f$, where $\bu_1, \dots, \bu_q$ are independent and
identically distributed (iid) uniformly in the unit cube $[0,1]^s$.
Obviously, $y_1, \dots, y_q$ each have the same probability distribution as $y = Q_{\bu} f$, where
$\bu$ is uniformly distributed in the unit cube $[0,1]^s$.
The moments of $\bar{y}$ about its mean can be found from the moments of $y$, using the well-known
properties of cumulants (see e.g. Sections 1.3 and 1.4 of Barndorff-Nielsen and Cox, 1989).
The $r$'th cumulant $\kappa_r(Y)$ of the random variable $Y$ is defined on p.6 of Barndorff-Nielsen and Cox (1989).
This has the following properties. For any number $a$, $\kappa_r(aY) = a^r \kappa_r(Y)$.
If $Y_1, \dots, Y_n$ are iid with the same probability distribution as $Y$ then
$\kappa_r(Y_1 + \dots + Y_n) = n \, \kappa_r(Y)$. Define the $r$'th moment about the mean
$\mu_r = E \big [ (Y - E(Y))^r \big ]$, $r=1,2,\dots$. The $\mu_r$'s can be expressed in terms
of the $\kappa_r$'s in the way described on p.7 of Barndorff-Nielsen and Cox (1989):
$\mu_2 = \kappa_2$, $\mu_3 = \kappa_3$, $\mu_4 = \kappa_4 + 3 \kappa_2^2$, etc..
Note that the random variable $y - I f$ has bounded support
since $|y - I f| \le \mathop{{\sum}'}_{\bh \in L^{\bot}} \, |\hat{f}(\bh)|$.
As is well-known, $E(\bar{y} - I f) = 0$. This follows immediately from the fact that
$E(y - I f) = 0$, which can be proved as follows.
\begin{align*}
E(y - I f) &= E(Q_u f - I f) \\
&= E \left(  \mathop{{\sum}'}_{\bh \in L^{\bot}} \, e^{2 \pi i \bh \cdot \bu} \, \hat{f}(\bh) \right) \\
&= \mathop{{\sum}'}_{\bh \in L^{\bot}} \, \hat{f}(\bh) \, E(e^{2 \pi i h_1 u_1}) \cdots E(e^{2 \pi i h_s u_s})
\ \ \ \text{by independence of } u_1, \dots, u_s\\
&= 0,
\end{align*}
since, for integer $h$ and $u \sim U[0,1]$,
\begin{equation}
\label{basic_expectation}
E(e^{2 \pi i h u}) = \int_0^1 e^{2 \pi i h u} \, du =
\begin{cases}
1 &\text{if } h = 0 \\
0 &\text{otherwise}
\end{cases}
\end{equation}

L'Ecuyer and Lumieux (2000, Proposition 4) (see also L'Ecuyer, Munger and Tuffin, 2010)
provide an expression for $E((\bar{y} - I f)^2)$ in terms of the
Fourier coefficients of $f$. We may derive this expression as follows. Note that
$E((\bar{y} - I f)^2) = \text{Var}(\bar{y}) = E((y - I f)^2)/q$. Since $\bar{y} - I f$ is a real
number, it is equal to its complex conjugate, so that
\begin{equation*}
\bar{y} - I f = \mathop{{\sum}'}_{\bk \in L^{\bot}} \, e^{- 2 \pi i \bk \cdot \bu} \, \hat{f}^*(\bk),
\end{equation*}
where $\hat{f}^*(\bk)$ denotes the complex conjugate of $\hat{f}(\bk)$. Thus
\begin{align*}
E((y - I f)^2)
& = E \left( \mathop{{\sum}'}_{\bh \in L^{\bot}} \, e^{2 \pi i \bh \cdot \bu} \, \hat{f}(\bh)
\mathop{{\sum}'}_{\bk \in L^{\bot}} \, e^{- 2 \pi i \bk \cdot \bu} \, \hat{f}^*(\bk)
\right) \\
&= \mathop{{\sum}'}_{\bh \in L^{\bot}} \, \mathop{{\sum}'}_{\bk \in L^{\bot}} \,
\hat{f}(\bh) \, \hat{f}^*(\bk) \,
E ( e^{2 \pi i (\bh - \bk) \cdot \bu} ) \\
&= \mathop{{\sum}'}_{\bh \in L^{\bot}} \, |\hat{f}(\bh)|^2, \ \ \ \ \ \ \ \text{by } \eqref{basic_expectation}.
\end{align*}

The same kind of argument can be used to find expressions for
higher-order moments of $\bar{y}$, in terms of the Fourier coefficients of $f$.
For example, $E((\bar{y} - I f)^3) = E((y - I f)^3)/q^2$ and
\begin{align}
E((y - I f)^3)
& = E \left( \mathop{{\sum}'}_{\bh \in L^{\bot}} \, e^{2 \pi i \bh \cdot \bu} \, \hat{f}(\bh)
\mathop{{\sum}'}_{\bk \in L^{\bot}} \, e^{ -2 \pi i \bk \cdot \bu} \, \hat{f}^*(\bk)
\mathop{{\sum}'}_{\bl \in L^{\bot}} \, e^{ -2 \pi i \bl \cdot \bu} \, \hat{f}^*(\bl)
\right) \notag \\
&= \mathop{{\sum}'}_{\bh \in L^{\bot}} \, \mathop{{\sum}'}_{\bk \in L^{\bot}} \,
\mathop{{\sum}'}_{\bl \in L^{\bot}} \,
\hat{f}(\bh) \, \hat{f}^*(\bk) \, \hat{f}^*(\bl)
E ( e^{2 \pi i (\bh - \bk - \bl) \cdot \bu} ) \notag \\
\label{3rd_moment_intermediate}
&= \sum_{\cal C} \hat{f}(\bh) \, \hat{f}^*(\bk) \, \hat{f}^*(\bl),
\end{align}
where ${\cal C} = \{(\bh,\bk,\bl): \bh \in L^{\bot}, \bh \ne \bzero, \bk \in L^{\bot}, \bk \ne \bzero,
\bl \in L^{\bot}, \bl \ne \bzero, \bh-\bk-\bl = \bzero \}$,
by \eqref{basic_expectation}. Using the fact that a lattice is closed under subtraction, we see
that \eqref{3rd_moment_intermediate} is equal to
\begin{equation*}
\mathop{{\sum}'}_{\bh \in L^{\bot}} \, \hat{f}(\bh) \, \left (
\mathop{{\sum}'}_{\bk \in L^{\bot}, \bk \ne \bh} \, \hat{f}^*(\bk) \, \hat{f}^*(\bh-\bk)
\right )
\end{equation*}

\end{document}